%% file: sc21-rse-hpc-challenges.tex
\documentclass[conference]{IEEEtran}

\usepackage[english]{babel}
\usepackage[utf8x]{inputenc}
\usepackage[T1]{fontenc}

\usepackage[colorlinks=true, allcolors=blue]{hyperref}

\title{(R)SE challenges in HPC}

\author{\IEEEauthorblockN{{Jonas Thies}\IEEEauthorrefmark{1},
{Melven Röhrig-Zöllner}\IEEEauthorrefmark{2}, and
{Achim Basermann}\IEEEauthorrefmark{2}}

\IEEEauthorblockA{\IEEEauthorrefmark{1} \textit{Delft High Performance Computing Center}, \textit{Delft University of Technology}, \\
j.thies@tudelft.nl}
\IEEEauthorblockA{\IEEEauthorrefmark{2} \textit{Institute for Software Technology}, \textit{German Aerospace Center (DLR)}, Cologne, Germany\\
melven.roehrig-zoellner@dlr.de, achim.basermann@dlr.de}
}

\begin{document}
\maketitle

\begin{abstract}
\input{abstract.tex}
\end{abstract}

\section{Introduction}
It is a common observation that in the field of high-performance computing (HPC)
scientists only slowly adopt new software engineering techniques that are already successful
in e.g., web development or commercial applications.
This was described from the software engineer's point of view in~\cite{Basili2008}. 
We approach the topic from the HPC engineer's point of view.
In our opinion, improved training of HPC developers is an important step,
but it needs to address specific challenges inherent to HPC software. In contrast
to other fields, HPC software always has the design goal of achieving high hardware
efficiency, which in turn ensures energy efficiency~\cite{Zwart2020} and makes extreme-scale
applications feasible in the first place.
In addition to training, tools for programming and testing must be adjusted to work
correctly in an HPC environment.
In this paper, we present key aspects of software design, testing and performance engineering
for HPC software.

\section{Challenges in HPC software development}
We identify three key challenges that seem to be invariant with respect to
circumstances such as the actual application, hardware or programming skills of the developers. 

\paragraph*{First challenge} The life-cycle of HPC hardware is significantly shorter than
that of HPC
software, while at the same time software must be tailored to the hardware in order to
achieve optimal performance. In the course of a decade the
supercomputer hardware evolves dramatically (e.g. from vector processors to clusters of
CPUs, from single to multi-core processors, from commodity hardware to graphics or tensor
processing units). In contrast, much of the code base in use is at least twenty or thirty years old, and
developing e.g. a new aerodynamics code for industrial use may take decades
even with a large team and modern software engineering technology.

\paragraph*{Second challenge} The number of possible code paths grows exponentially in order
to provide
high performance.
A user's call to a simple basic linear algebra subroutine (BLAS)
may trigger any of dozens of implementations, differing in arithmetic (real, complex),
precision (half/single/double/quad, or vendor-specific variants thereof),
data layout (e.g. row- or column major matrix storage), threading mechanisms or GPU programming model,
SIMD hardware (SSE/AVX/ARM/...).
This leads to an explosion of combinations of (possibly generated) code paths.
In some cases the testing responsibility is with hardware-specific vendor libraries (like the Intel MKL or CUBLAS),
but `hand-optimized' code for special purposes must still be tested efficiently and comprehensively.

\paragraph*{Third challenge} It is difficult to reproduce performance results.
Due to the fast pace at which the hardware develops, another user of a code or algorithm
may not have a comparable machine in terms of speed, memory, parallelism, or even architecture.
Simple and general machine models allow
 assessing the efficiency of an implementation across platforms, as we will discuss in
Section~\ref{sec:perf_port}. They can also ensure that the system's hardware and software
are configured appropriately as even small changes can reduce the performance by a factor of
two or more.

\section{Designing HPC software}
In order to meet the challenge of the mismatched software/hardware life-cycle it is crucial
to achieve \emph{separation of concerns} in HPC applications. The climate scientist who
develops a new model component, or the numerical mathematician who develops a new algorithm,
cannot port the software to the next few generations of hardware in the life-cycle
of the code. Instead, they need robust interfaces through which the application,
algorithms and low-level implementations (kernels) are separated.
For decades, the libraries BLAS and LAPACK~\cite{LAPACK} provide a commonly used interface
to linear algebra building blocks.
However, the choice of the granularity of the \emph{building blocks} as well as their interfaces
are \emph{architectural decisions}.
In particular, to obtain high efficiency, one needs to optimize the node-level performance as well as 
the communication. Both of these optimizations often affect the code globally e.g.,
through the memory-layout and the distribution of data.
So new advances such as communication-avoiding algorithms (or better: data-transfer avoiding algorithms, see~\cite{Demmel2012})
cannot always be implemented just under the hood, see~\cite{Springer2018,RoehrigZoellner2021preprint} for examples.

In~\cite{Thies2016lncs} we described a layered software
architecture for a sparse eigenvalue solver library with applications in quantum physics.
The kernel interface we proposed (see also the PHIST software,~\cite{Thies2020phist}) allows
the algorithms and applications layers to work with multiple backends, among which are large
open source libraries optimized for portability (e.g. Trilinos) and hand-optimized
hardware-specific ones like GHOST~\cite{Kreutzer2016ghost}. PHIST provides both unit tests
for the backends and performance models for all operations used in its algorithms. That way,
a new development on the hardware side can be met by either the extension of an existing
implementation or a completely new one, and the new component can be readily tested in terms
of correctness and performance. The algorithms and applications layers only have to be
modified or extended if new needs arise on their respective level. The significantly larger
HPC software project Trilinos~\cite{trilinos-website} takes the approach of offering a large
number of interoperable `packages' which may have different life cycles. While this also
results in a manageable overall software, it may incur smaller or larger interface
adaptations for users from time to time. The package concept is taken to the next level by the xSDK
project (\url{https://xsdk.info}), which aims at gradually improving the software quality
and interoperability of a whole landscape of HPC libraries and applications by defining
common rules and recommendations.

\section{Testing}
Above, we mentioned the potentially large amount of (generated) code that needs to be covered by unit testing.
In addition, HPC software often employs multiple parallelization levels at once
(e.g., OpenMP for CPU multi-threading, MPI for communication between nodes and CUDA for GPUs).
This can lead to functionality that is available but not well-tested.
We propose to anticipate typical bugs in HPC codes and to design unit tests specifically to trigger them
(similar to \emph{white-box testing} but with multiple different possible implementations in mind).
In PHIST, for instance, all basic linear algebra tests are executed for aligned and unaligned memory cases
to locate invalid use of SIMD operations. Other typical `parallel bugs' include race
conditions and deadlocks.
Beyond such HPC-specific tests, one needs to explore the space of available combinations of
hardware features with a finite test-matrix
by selecting a hopefully representative subset.

A practical problem is that test frameworks typically lack support for MPI applications,
as well as for other parallelization techniques such as OpenMP or CUDA.
At least MPI support is crucial to run the tests on current supercomputers.
An exception is pFUnit~\cite{pFUnit} for Fortran which supports MPI and OpenMP.
For C++ we provide an extended version of GoogleTest with MPI support at \url{https://github.com/DLR-SC/googletest_mpi}.
It features correct I/O and handling of test results in parallel, as well as collective
assertions.

\section{Performance Portability}\label{sec:perf_port}
In many papers, performance results are reported in terms of `scalability' of a
parallel program: either the speed-up achieved by using more processes to solve the same
problem (strong scalability), or the parallel efficiency when increasing the problem size
with the number of processes (weak scalability).
Such results are not necessarily helpful for comparing the performance on different machines.
A better way is to identify the bottleneck in the computation and to report resource utilization
with respect to that bottleneck: In the vast majority of HPC codes, the bottleneck is either
floating point arithmetic (`compute bound' applications), or data movement (`memory bound' or `communication bound').
This allows estimating the attainable performance by the roofline performance
model~\cite{Williams2009}.
With some measurements of cache/memory/network bandwidths and counting of operations and data volumes,
one can calculate the achieved performance relative to the (modeled) attainable performance.
This relative roofline performance provides a criterion that is independent of the underlying hardware.

Unfortunately, tools cannot easily compute this automatically as it requires high-level insight into the algorithms.
For instance, an unfavorable memory access pattern may or may not be avoidable by code or
algorithm restructuring. A performance model can be formulated to predict the optimal runtime of the bad access pattern
(labelling a good implementation as efficient).
Alternatively, a model can predict the runtime of the actual amount of data traffic needed to perform the operation in an ideal setting
(highlighting this part of the algorithm as inefficient).
We therefore decided to build the roofline model manually into the timing functionality 
for all basic operations of the PHIST software, giving the user a choice of these two
variants (realistic vs. idealized)~\cite{Thies2016lncs,Thies2020phist}. When running the
same application on two different machines, one can then compare the overall roofline performance,
or the performance achieved by individual operations, even for different hardware and/or backends.

\section{Summary}
In this overview of software engineering challenges specific to HPC, we argued that HPC
applications are particularly vulnerable to poor software engineering because their
development and use typically outlasts several generations of HPC hardware.
Basic functionality needs to be implemented `close to the hardware',
so that supporting (combinations of) multiple architectures and programming models
leads to additional complexity and to a large amount of (generated) code which has to be tested.
And finally, as the hardware develops rapidly, it is difficult to compare performance results on different machines and hardware architectures.

We illustrated some aspects of software design, unit testing and of the portability of performance results
 with a practical solution from our own field of research, (sparse) linear algebra.
The points we would like to highlight are \emph{separation of concerns} when designing the software,
\emph{anticipating HPC-specific bugs}, and using \emph{performance models} to validate the efficiency of an implementation
across different hardware.

\bibliographystyle{IEEEtran}
\bibliography{IEEEabrv,sc21-rse-hpc-challenges}

\end{document}

%% file: abstract.tex
We discuss some specific software engineering challenges in the field of
high-performance computing, and argue that the slow adoption of SE tools and techniques
is at least in part caused by the fact that these do not address the HPC challenges
`out-of-the-box'. By giving some examples of solutions for designing, testing and
benchmarking HPC software, we intend to bring software engineering and HPC closer together.